\documentclass{article}    

\usepackage{graphicx}

\title{\textbf{Hamiltonian Based nRules \linebreak{} -- Time's Arrow}}  
\author{Richard Mould\footnote{Department of Physics and Astronomy, State University of New York, Stony Brook,
\mbox{New York} 11794-3800; http://ms.cc.sunysb.edu/\~{}rmould}}  
\date{}    
   
\begin{document}             

\maketitle              

\begin{abstract}

The auxiliary rules of quantum mechanics can be written without the Born rule by using what are called the nRules.  The
nRules are understood in part by making certain modifications in the Hamiltonian.  In this paper, those modifications
are written directly into the nRules, reducing their number from four to three.  It is shown that the nRules in either
form provide for a definite direction in time, guaranteeing that a statistically irreversible interaction is
absolutely irreversible.

\end{abstract}

\section*{Introduction}
	It is shown in another papers \cite{RM1, RM2} that the Born rule (relating probability with square modulus) and the
other auxiliary rules of standard quantum mechanics are disposable.  In their place two other rule-sets have been
proposed (called the nRules and the oRules) that introduce probability into quantum mechanics through probability
current alone.  There is no attempt to explain these rules.  The strategy has been to show that these two rule-sets
work over a wide range of examples without attempting a theoretical justification.   We assume that a theory will one
day be found to cover the auxiliary rules of quantum mechanics, so it is important to have the right rules.    

	The Schr\"{o}dinger solutions are continuous in all variables for the most part; but occasionally, there is a
discontinuity in one or more variables (except time) that leads to a \emph{quantum jump}.  An example is the change that
takes place when an electron drops from a higher atomic orbit to a lower one, discontinuously emitting a photon in the
process. When a system encounters a quantum jump of this kind, the Hamiltonian in standard quantum mechanics is $H =
H_0 + H_{01} + H_1$, where $H_0$ drives the original system $S_0$, $H_1$ drives the final system $S_1$, and $H_{01}$
drives the interaction between the two.  The separation of the Hamiltonian into these parts is possible because the
variables are indexed to distinguish systems $S_0$ and $S_1$.  However, a truncated Hamiltonian equal to $H = H_0 +
H_{01}$ was found in Ref.\ 1 to be required by the nRules.  This has the desired effect of `bridging' the
discontinuous quantum gap; and at the same time, forbidding any further evolution until there has been a stochastic
hit on the original system.  When that happens, the system acquires new boundary conditions that launch the new state
$S_1$.  This abrupt change of state (from $S_0$ to $S_1$) is the ``collapse" of the wave function.   

Adopting the truncated Hamiltonian enforces nRule (4) -- the fourth of the four nRules; so if the Hamiltonian were
directly referred to in the auxiliary nRules, then  nRule (4) would be unnecessary.  When this is done (below), the
total number of nRules is reduced from four to three.  Where a distinction between the new three-rule rule-set and the
previous four-rule rule-set is desirable, they will be designated nRules$^3$ and nRules$^4$ respectively.

Finally, it is shown that the nRules$^3$ exhibited below insure that probability current can only flow from the low
entropy side of an irreversible interaction to the high entropy side.  Thermodynamics claims that a reverse process is
very improbable, but the nRules$^3$ \emph{insure} that a current reversal of this kind will not occur across any
irreversible quantum gap.  Therefore, they guarantee the omni-direction of time's arrow.  The same guarantee is shown
(below) to follow from the original nRules$^4$ that are developed in previous papers.

\section*{The nRules$^3$}
They are:

\vspace{.4cm}

\noindent
\textbf{nRule$^3$ (1)}: \emph{If an irreversible interaction connects  complete components that are discontinuous with
each other in some variable, then the high entropy component will not be driven by its own Hamiltonian.}

\hangindent=.2in
[\textbf{note:} \emph{Complete components} contain all of the symmetrized objects in the universe.  Each included
object is itself complete in that it is not a partial expansion in some representation.  The entropy change across a
discontinuous gap is therefore the entropy change of the entire universe.]

\hangindent=.2in
[\textbf{note:} A complete component's \emph{own Hamiltonian} is that part of the Hamiltonian whose variables are
exclusively those of the component.  These variables only intermingle with the variables of another complete component
in the `interaction' part of the Hamiltonian.]

\vspace{.4cm}

\noindent
\textbf{nRule$^3$ (2)}: \emph{For a system of total square modulus s that has n launch components, a stochastic trigger
will choose stochastically from among them.  The probability per unit time of such a choice among m of these components
at time t is given by  $(\Sigma_mJ_m)/s$, , where the square modular current $J_m$ flowing into the $m^{th}$  component at
that time is positive.}

\hangindent=.2in
\textbf{note:} A \emph{launch component} is the high entropy component of an irreversible and discontinuous gap.

\vspace{.4cm}

\noindent
\textbf{nRule$^3$ (3)}: \emph{If a launch component is stochastically chosen, it will be driven by the entire
Hamiltonian of the new solution.  All other components in the original superposition will be immediately reduced to zero.}

\section*{Time's Arrow}
	Let probability current flow across the discontinuous gap between components $C_0$ and $\underline{C}_1$ in the first
row of Eq.\ 1, where component $C_0$ has a lower entropy than $\underline{C}_1$. 
\begin{eqnarray}
\Phi(t_{sc} >t \ge t_0) &=& C_0(t_0) \rightarrow C_0(t_{sc} >t \ge t_0) + \underline{C}_1(t_{sc} >t \ge t_0) \\
\Phi(t \ge t_{sc}) &=& C_1(t = t_{sc}) \rightarrow C_1(t > t_{sc})\nonumber
\end{eqnarray}
The first row represents the time between $t_0$ and the time $t_{sc}$ of a stochastic hit, and the second row is the
collapsed wave function after a stochastic hit on $C_1$.  The higher of two entropy components is underlined in this
treatment.  The arrow represents a continuous evolution and the + sign is a discontinuous evolution in some variable. 
According to thermodynamics, the direction of flow in Eq.\ 1 will \emph{most probably} be from left to right. 

Now consider how this equation would look if the direction of the current were reversed.  First, suspend nRule$^3$ (1) to
give
\begin{equation}
\Phi(t \ge t_0) = C_0(t_{sc} > t) \leftarrow C_0(t > t_0) + \underline{C}_1(t \ge t_0) \leftarrow \underline{C}_1(t_0) 
\end{equation}
Both $\underline{C}_1$ components in this equation are the same component at different times.  In that case there is no
stochastic hit because there is no probability current flowing into $\underline{C}_1$ as required by nRule$^3$ (2).  If
we now restore Rule$^3$ (1), then the continuous evolution $\underline{C}_1(t > t_0) \leftarrow \underline{C}_1(t_0)$
cannot occur because the component will not be driven by its own Hamiltonian; so Eq.\ 2 cannot occur.  This means that
probability current can only flow from left to right as shown in Eq.\ 1, going from the lower to the higher entropy
across the discontinuous gap.  Thermodynamics says that a reverse flow  is very improbable, but the
nRules$^3$ say it is impossible.  An unambiguous direction of time is thereby established in the direction of higher
entropy in a quantum mechanical system in which stochastic choices occur. 

	If the system is governed by the auxiliary nRules$^4$ found in Refs.\ 1 and 2, its forward evolution will be the same
as that given in Eq.\ 1, where the underlined components are now called `ready' components.  The wording of nRule$^4$ (1)
is not optimal for our present purpose because it contains the word ``predecessor" or ``initial".  However, the high and
low entropy language of nRule$^3$ (1) can easily be substituted.

The backward evolution of the system is again described in Eq.\ 2 when nRule$^4$ (4) is suspended.  As before, both
$\underline{C}_1$ components are the same component at different times.  If nRule$^4$ (4) is restored, then Eq.\ 2 cannot
exist because current cannot flow from $\underline{C}_1(t_0)$ to $\underline{C}_1(t > t_0)$, or from $\underline{C}_1(t
> t_0)$ across the gap to $C_0(t > t_0)$.  Therefore, only the forward flow of current in Eq.\ 1 is possible.  Again, an
unambiguous direction of time is established in the direction of higher entropy in a quantum mechanical system in which
stochastic choices occur.

\end{document}